\documentclass{ws-ijmpa}

\usepackage{bm}
\usepackage{amsmath}
\usepackage{amsfonts,amsbsy,amsxtra}
\usepackage{amssymb,latexsym}
\usepackage{color}
\usepackage{dcolumn}
\usepackage{textcomp}
\usepackage{booktabs}
\usepackage{soul}
\usepackage{lineno}

\newcommand{\figref}[1]{{Fig.~\ref{fig:#1}}}
\newcommand{\Figref}[1]{{Figure~\ref{fig:#1}}}

\newcommand{\others}{\textit{et al.}}

\newcommand{\antibar}[1]{%
  \overline{#1}%
}

\renewcommand{\Pr}{\ensuremath{\rho}}
\newcommand{\Prthree}{\ensuremath{\rho_3}}

\newcommand{\Paone}{\ensuremath{a_1}}
\newcommand{\Patwo}{\ensuremath{a_2}}

\newcommand{\Pfzero}{\ensuremath{f_0}}

\newcommand{\Pftwo}{\ensuremath{f_2}}

\newcommand{\Ppi}{\ensuremath{\pi}}
\newcommand{\Ppione}{\ensuremath{\pi_1}}
\newcommand{\Ppitwo}{\ensuremath{\pi_2}}

\newcommand{\Ppip}{\ensuremath{\pi^+}}
\newcommand{\Ppim}{\ensuremath{\pi^-}}

\newcommand{\PKm}{\ensuremath{K^-}}

\newcommand{\Pqq}{\ensuremath{q}}
\newcommand{\Pqqbar}{\ensuremath{\antibar{q}}}

\newcommand{\PPb}{\ensuremath{\text{Pb}}}
\newcommand{\twopion}{\ensuremath{\Ppip\Ppim}}
\newcommand{\threepion}{\ensuremath{\Ppim\Ppip\Ppim}}

\newcommand{\abs}[1]{{|{#1}|}}

\newcommand{\measresult}[4]{%
  \ensuremath{#1%
    \ifthenelse{\equal{#2}{}}%
    {}%
    {\pm #2%
      \ifthenelse{\equal{#3}{}}%
      {}%
      {_\text{stat.}}%
    }%
    \ifthenelse{\equal{#3}{}}%
    {}%
    {\pm #3_\text{syst.}}\text{#4}%
  }%
}

\newcommand{\jpc}{\ensuremath{J^{PC}}}

\newcommand{\wavespec}[7]{\ensuremath{#1^{#2#3}}\!\!
  \ensuremath{#4^{#5}}\!\! \ensuremath{[#6] #7}} % J, P, C, M, reflectivity, isobar, L

\newcommand{\gevc}{~\ensuremath{\text{GeV}\! / c}}
\newcommand{\gevcsq}{~\ensuremath{(\text{GeV}\! / c)^2}}
\newcommand{\mevcc}{~\ensuremath{\text{MeV}\! / c^2}}
\newcommand{\gevcc}{~\ensuremath{\text{GeV}\! / c^2}}
\newcommand{\tenpow}[2][]{%
  \ifthenelse{\equal{#1}{}}
  {\ensuremath{10^{#2}}}
  {\ensuremath{{#1} \cdot 10^{#2}}}
}

\begin{document}

\markboth{Boris Grube}
{Diffractive Dissociation of 190~\gevc\ \Ppim\ at COMPASS}

\catchline{}{}{}{}{}
%
%%%%%%%%%%%%%%%%%%%%%%%%%%%%%%%%%%%%%%%%%%%%%%%%%%%%%%%%%%%%%%%%%%%%

\title{%
  DIFFRACTIVE DISSOCIATION OF 190~\gevc\ \Ppim\ \\
  INTO \threepion\ FINAL STATES AT COMPASS
}

\author{BORIS GRUBE for the COMPASS Collaboration}

\address{%
  Excellence Cluster Universe, Technische Universit\"at M\"unchen, \\
  Boltzmannstr. 2, 85748 Garching, Germany. \\
  bgrube@ph.tum.de}

\maketitle

\begin{history}
\received{Day Month Year}
\revised{Day Month Year}
\end{history}

\begin{abstract}
  %COMPASS is a multi-purpose fixed-target experiment at the CERN Super
  %Proton Synchrotron investigating the structure and spectrum of
  %hadrons.
  We present results from a Partial-Wave Analysis (PWA) of diffractive
  dissociation of 190~\gevc\ \Ppim\ into \threepion\ final states on
  nuclear targets.
  %This reaction provides clean access to the
  %light-quark meson spectrum up to 2.5~\gevcc.
  A PWA of the data sample taken during a COMPASS pilot run in 2004 on
  a \PPb\ target showed a significant spin-exotic $\jpc = 1^{-+}$
  resonance consistent with the controversial $\Ppione(1600)$, which
  is considered to be a candidate for a non-\Pqq\Pqqbar\ mesonic
  state. In 2008 COMPASS collected a large diffractive \threepion\
  data sample using a hydrogen target. A first comparison with the
  2004 data shows a strong target dependence of the production
  strength of states with spin projections $M = 0$ and $1$.

\end{abstract}

\ccode{PACS numbers: 13.25.-k, 13.85.-t, 14.40.Be, 29.30.-h}

\vspace{2ex}

The COmmon Muon and Proton Apparatus for Structure and Spectroscopy
(COMPASS)\cite{compass} is a fixed-target experiment at the CERN
Super Proton Synchrotron. It is a two-stage spectrometer that covers a
wide range of scattering angles and particle momenta with high angular
resolution.
%and is equipped with hadronic and electromagnetic
%calorimeters so that COMPASS can reconstruct final states with charged
%as well as neutral particles.
The target is surrounded by a Recoil
Proton Detector (RPD) that measures the time of flight of the recoil
protons. COMPASS uses the M2 beamline which can deliver secondary
hadron beams with a momentum of up to 300\gevc\ and a maximum
intensity of \tenpow[5]{7}~$\text{sec}^{-1}$. The negative hadron beam
consists of 96.0~\% \Ppim and 3.5~\% \PKm. Two ChErenkov Differential
counters with Achromatic Ring focus (CEDAR) upstream of the target are
used to identify the incoming beam particles.

During a pilot run in 2004 and subsequent data taking periods in 2008
and 2009 COMPASS has acquired large data sets of diffractive
dissociation of 190\gevc\ \Ppim\ on H$_2$, Ni, W, and \PPb\
targets. In these events the beam pion is excited to some resonance
$X^-$ via $t$-channel Reggeon exchange with the target. At 190\gevc\
the process is dominated by Pomeron exchange.
%so that
%isospin and $G$-parity of the intermediate state $X^-$ are that of the
%beam particle.
Diffractive reactions are known to exhibit a rich spectrum of produced
states and are characterized by two kinematic variables: the square of
the total center-of-mass energy and the squared four-momentum transfer
from the incoming beam particle to the target, $t = (p_\text{beam} -
p_X)^2$. It is customary to use the variable $t' = \abs{t} -
\abs{t}_\text{min}$ instead of $t$, where $\abs{t}_\text{min}$ is the
minimum value of $\abs{t}$ allowed by kinematics.

\begin{figure}[t]
  \begin{center}
    \begin{minipage}[t]{0.44\textwidth}
      \includegraphics[width=\textwidth]{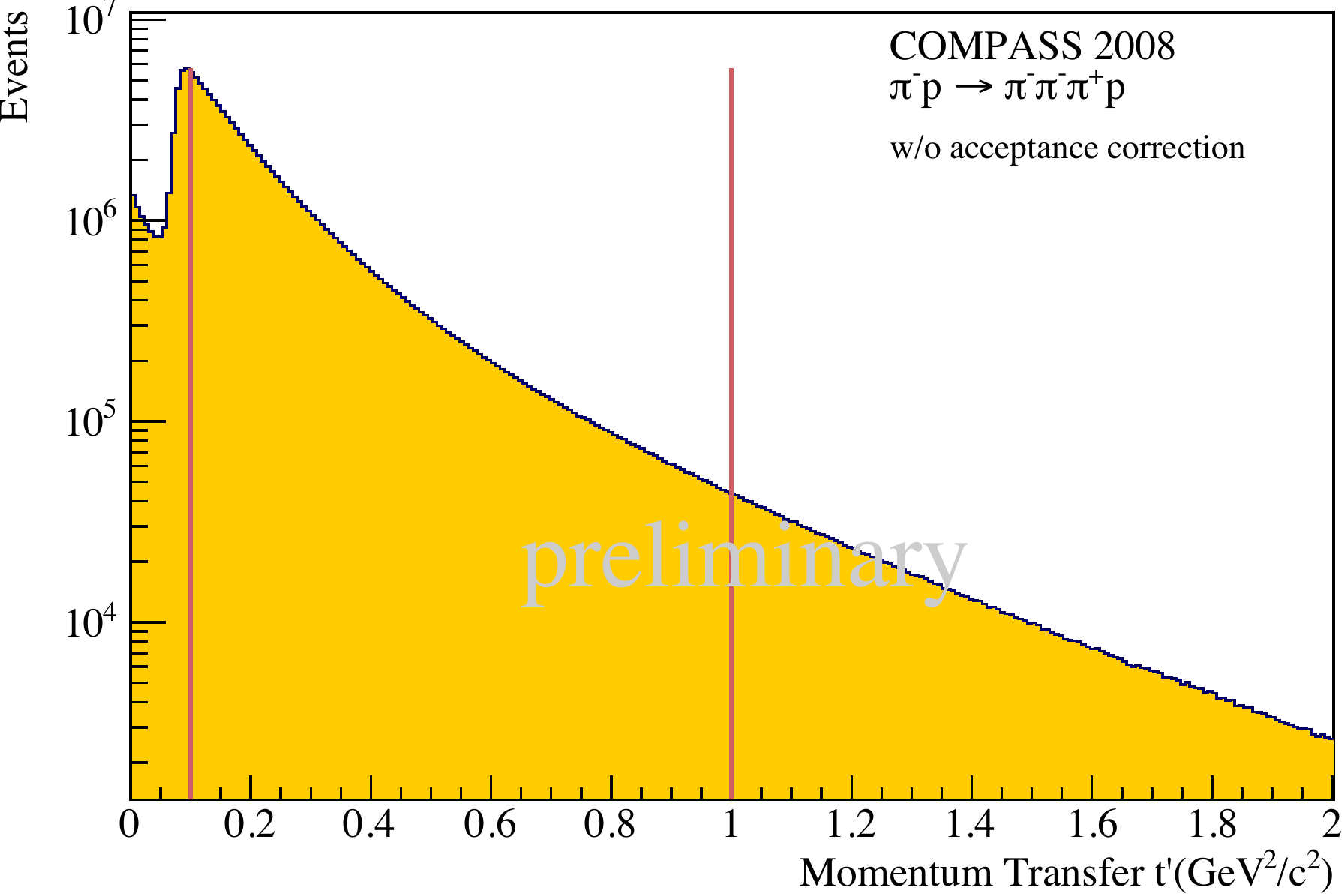}
      \caption{Squared four-momentum transfer $t' = \abs{t} -
        \abs{t}_\text{min}$. The analyzed region is indicated by the
        vertical lines}
      \label{fig:tPrime}
    \end{minipage} \hfill
    \begin{minipage}[t]{0.44\textwidth}
      \includegraphics[width=\textwidth]{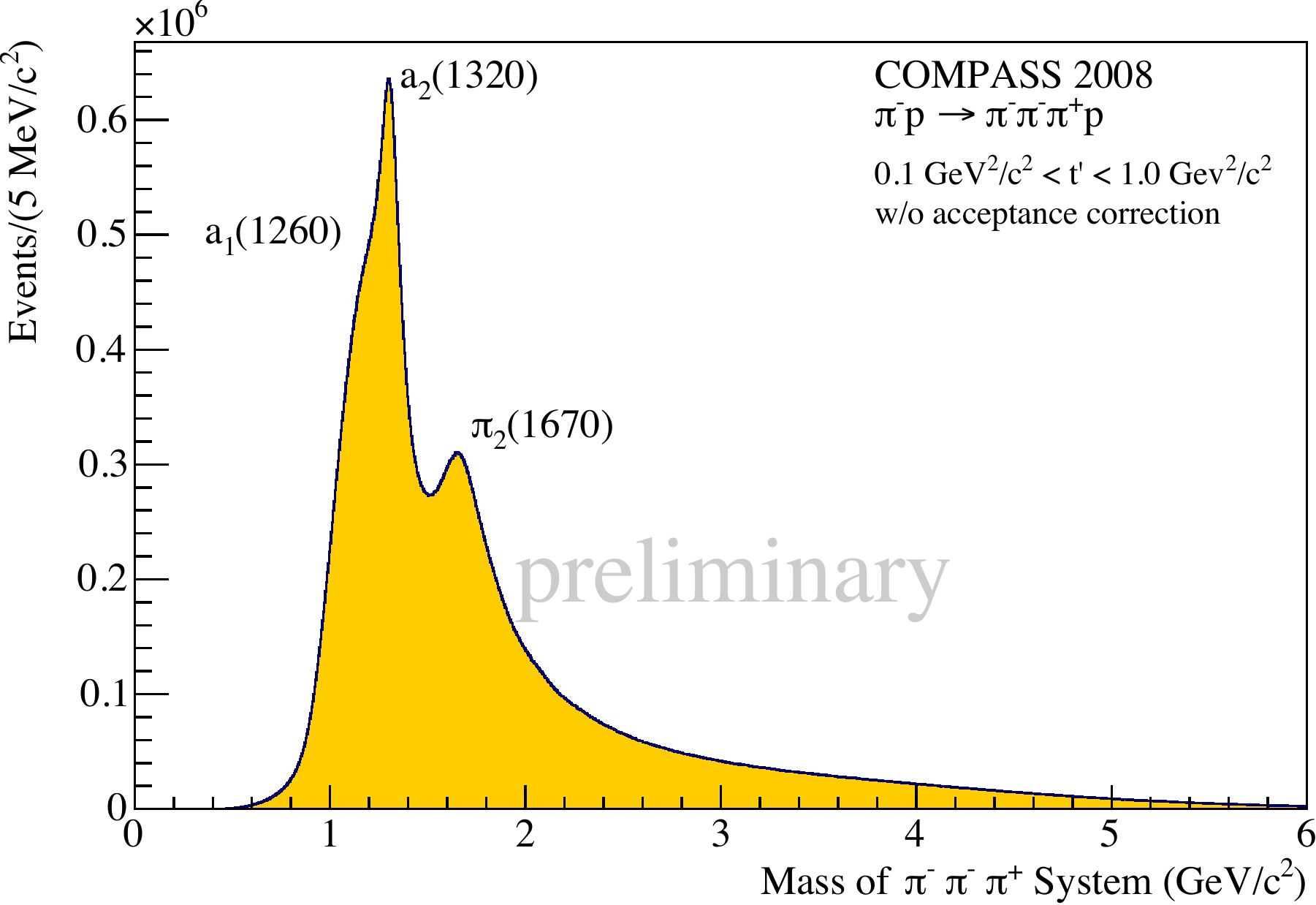}
      \caption{\threepion\ invariant mass distribution of the selected
        data sample for $t' \in [0.1, 1.0]\gevcsq$.}
      \label{fig:threePiMass}
    \end{minipage}
  \end{center}
\end{figure}

In 2004 the trigger selected one incoming and at least two outgoing
charged particles, whereas in 2008 a signal from the recoil proton was
required in the RPD. In the offline event selection diffractive events
were enriched by an exclusivity cut of $\pm 4$~GeV around the nominal
beam energy. The $t'$ region between 0.1 and 1.0\gevcsq\ was selected
for the analysis (see \figref{tPrime}).
%About 420\,000 (2004,
%Pb target) and xx~events (2008, H$_2$ target), respectively, in the
%mass range between 0.5 and 2.5\gevcc\ pass the selection cuts.

\Figref{threePiMass} shows the \threepion\ invariant mass distribution
of the selected 2008 data sample. It exhibits clear structures in the
mass regions of the well-known resonances $\Paone(1260)$,
$\Patwo(1320)$, and $\Ppitwo(1670)$. In order to find and disentangle
the various resonances in the data, a PWA was performed, in which the
total cross section was assumed to factorize into a resonance and a
recoil vertex. The isobar model\cite{isobar} is used to decompose the
decay $X^- \to \threepion$ into a chain of successive two-body decays:
The $X^-$ with quantum numbers \jpc\ and spin projection $M^\epsilon$
decays into a di-pion resonance, the so-called isobar, and a bachelor
pion. The isobar has spin $S$ and a relative orbital angular momentum
$L$ with respect to $\Ppim_\text{bachelor}$. A partial wave is thus
defined by $\jpc M^\epsilon[\text{isobar}]L$, where $\epsilon = \pm 1$
is the reflectivity\cite{reflectivity}.
%The $X^-$ decay amplitudes are calculated in the
%Gottfried-Jackson frame using non-relativistic Zemach
%tensors\cite{zemach} and do not contain any free parameters.
The production amplitudes are determined by extended maximum
likelihood fits performed in 40\mevcc\ wide bins of the three-pion
invariant mass $m_X$. In these fits no assumption is made on the
produced resonances $X^-$ other then that their production strengths
are constant within a $m_X$ bin. The PWA model includes five \twopion\
isobars\cite{compassExotic}: $\Ppi\Ppi$ $s$-wave, $\Pr(770)$,
$\Pfzero(980)$, $\Pftwo(1270)$, and $\Prthree(1690)$.
%were described using relativistic Breit-Wigner line
%shape functions including Blatt-Weisskopf barrier penetration
%factors. For the \twopion\ $S$-wave we use the parameterization
%from\cite{vesSigma} with the $\Pfzero(980)$ subtracted from the
%elastic \Ppi\Ppi\ amplitude and added as a separate Breit-Wigner
%resonance.
It consists of 41~partial waves with $J \leq 4$ and $M \leq 1$ plus
one incoherent isotropic background wave. In order to describe the
data, mostly positive reflectivity waves are needed. This corresponds
to production with natural parity exchange.

\begin{figure}[t]
  \begin{center}
    \begin{minipage}[t]{0.44\textwidth}
      \includegraphics[width=\textwidth]{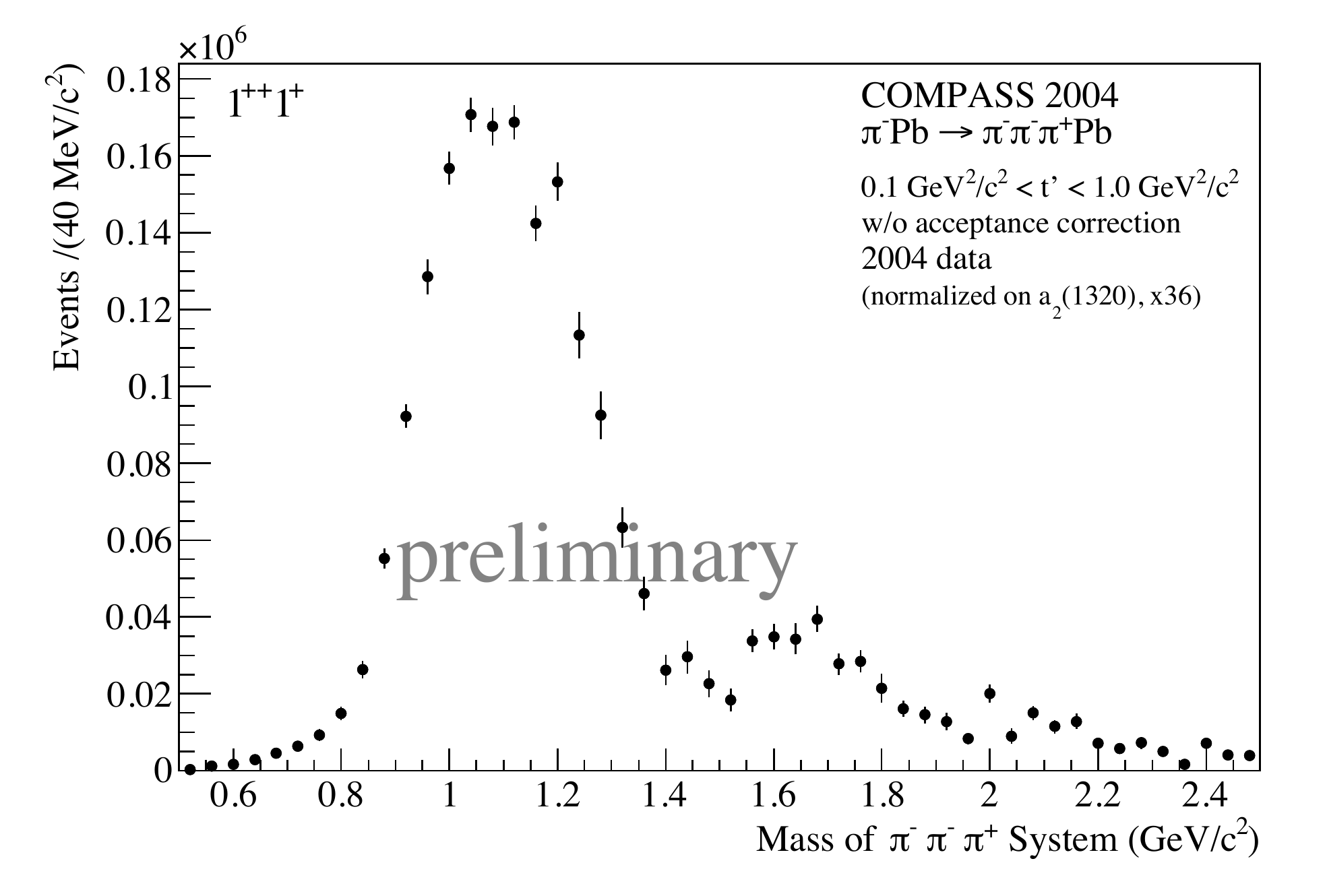} \\
      \includegraphics[width=\textwidth]{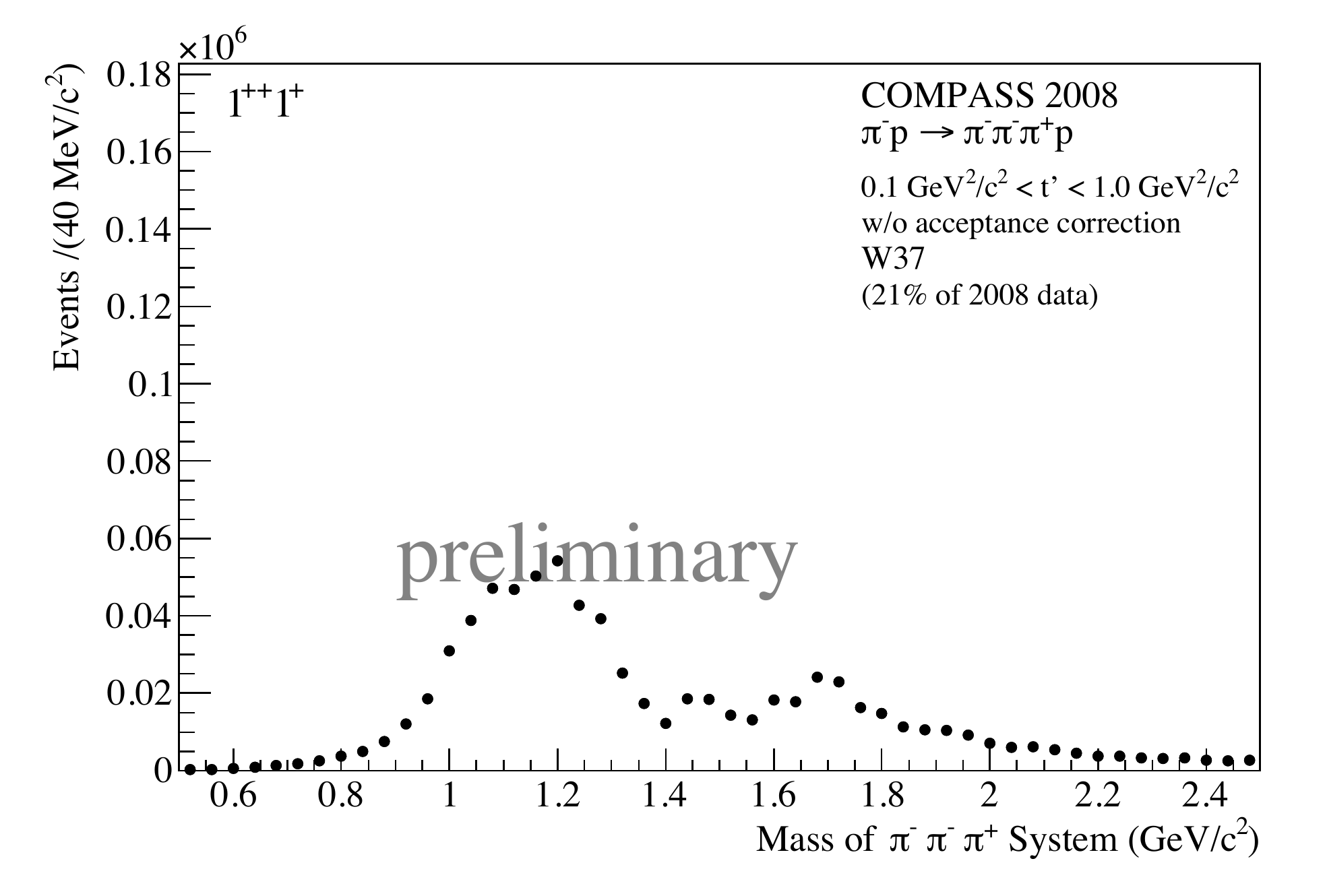}
    \end{minipage} \hfill
    \begin{minipage}[t]{0.44\textwidth}
      \includegraphics[width=\textwidth]{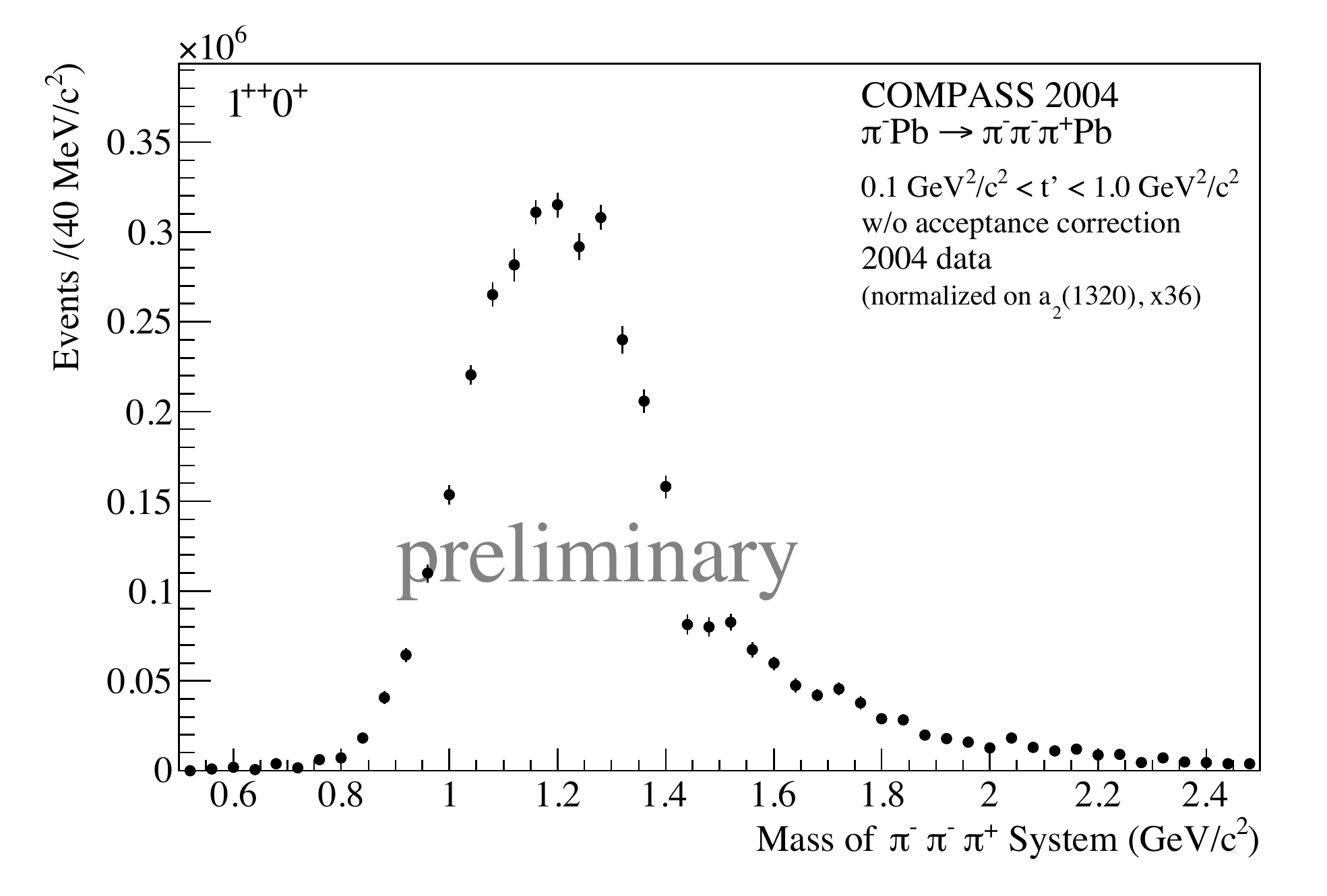} \\
      \includegraphics[width=\textwidth]{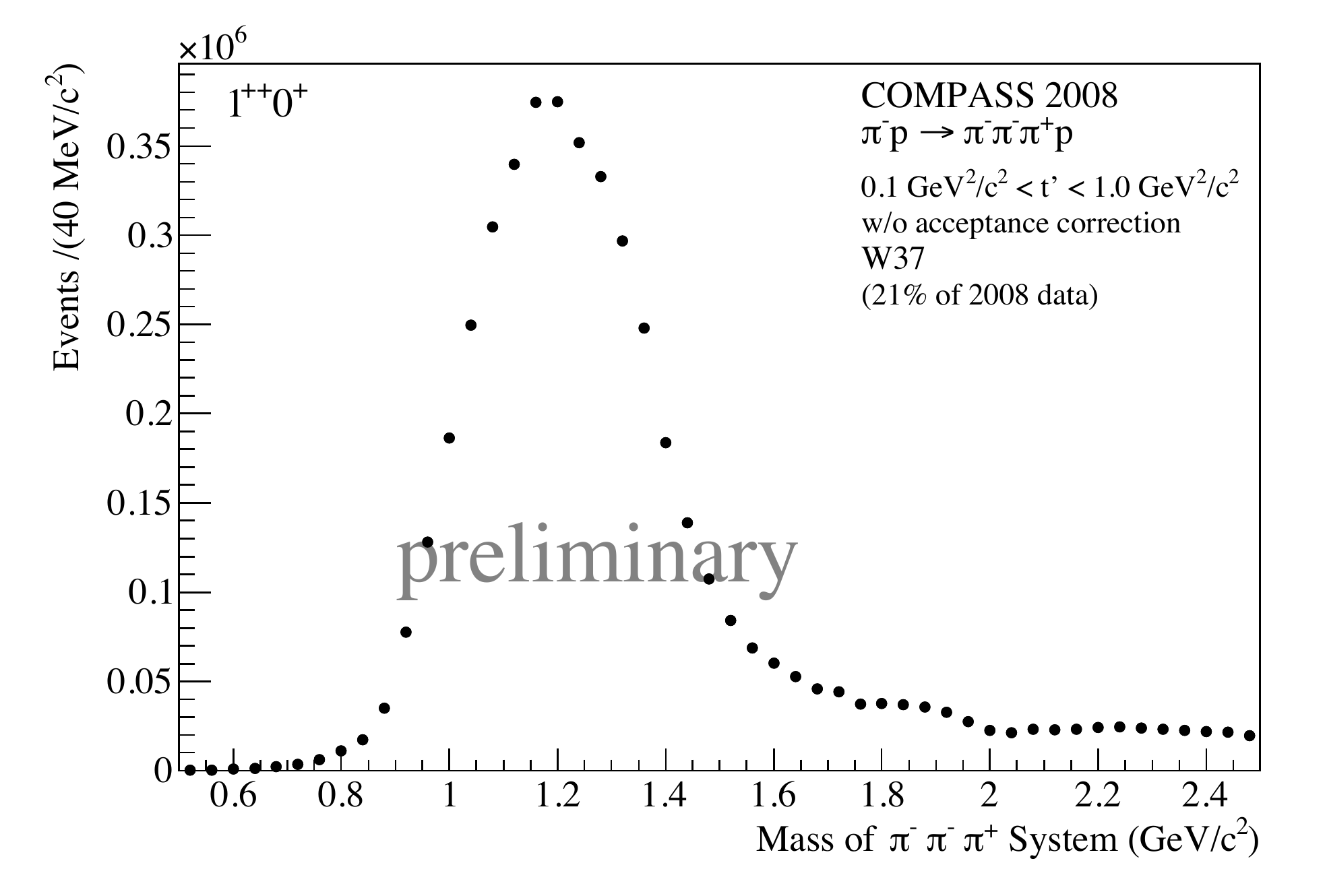}
    \end{minipage}
  \end{center}
  \caption{Normalized intensity sums of the $\jpc = 1^{++}$ partial
    waves for different spin projection quantum numbers $M = 1$ on the
    left and $M = 0$ on the right hand side. The top row shows data
    from the Pb, the bottom row data from the H$_2$ target. The wave
    intensities are dominated by a broad structure around 1.2\gevcc\
    which is the $\Paone(1260)$.}
  \label{fig:MDep}
\end{figure}

The three most dominant waves \wavespec{1}{+}{+}{0}{+}{\Pr\Ppi}{S},
\wavespec{2}{+}{+}{1}{+}{\Pr\Ppi}{D}, and
\wavespec{2}{-}{+}{0}{+}{\Pftwo\Ppi}{S} contain resonant structures
that correspond to the $\Paone(1260)$, $\Patwo(1320)$, and
$\Ppitwo(1670)$, respectively. The resonance parameters extracted from
the 2004 data are in good agreement with the PDG
values\cite{compassExotic}. In addition the 2004 data exhibit a
resonant peak around 1660\mevcc\ in the spin-exotic
\wavespec{1}{-}{+}{1}{+}{\Pr\Ppi}{P} wave consistent with the disputed
$\Ppione(1600)$\cite{compassExotic}. A first comparison of the 2008
H$_2$ data with the 2004 Pb data without acceptance corrections shows
a surprisingly large dependence on the target material. The data ---
normalized to the narrow $a_2(1320)$ resonance in the
\wavespec{2}{+}{+}{1}{+}{\Pr\Ppi}{D} wave --- exhibit a strong
suppresssion of $M = 1$ waves on the H$_2$ target, whereas the
corresponding $M = 0$ waves are enhanced such that the intensity sum
over $M$ remains about the same. As an example \figref{MDep} shows
this effect for the $a_1(1260)$ peak in the $\jpc = 1^{++}$ waves.

\section*{Acknowledgments}

This work is supported by the German
%Bundesministerium f\"ur Bildung und Forschung
BMBF, the Maier-Leibnitz-Labor der LMU und TU M\"unchen, the
DFG Cluster of Excellence \emph{Origin and Structure of the Universe},
and CERN-RFBR grant 08-02-91009.

\end{document}